\documentclass[amsmath,amssymb,prd,notitlepage]{revtex4-1}

\usepackage[margin=1.0in]{geometry}

\def\al{\alpha}
\def\be{\beta}
\def\ga{\gamma}

\def\la{\lambda}

\def\ph{\phi}

\def\ch{\chi}

\def\om{\omega}

\def\De{\Delta}

\def\La{\Lambda}

\def\Ps{\Psi}
\def\Om{\Omega}

\def\cL{{\cal L}}

\def\half{{\textstyle{1\over 2}}}
\def\ol{\overline}
\def\prt{\partial}

\def\ket#1{|{#1}\rangle}
\def\po{P^0}
\def\opo{\ol{P}{}^0}

\def\Re{\hbox{Re}\,}
\def\Im{\hbox{Im}\,}

\def\lsim{\mathrel{\rlap{\lower4pt\hbox{\hskip1pt$\sim$}}
    \raise1pt\hbox{$<$}}}
\def\gsim{\mathrel{\rlap{\lower4pt\hbox{\hskip1pt$\sim$}}
    \raise1pt\hbox{$>$}}}

\newcommand{\beq}{\begin{equation}}
\newcommand{\eeq}{\end{equation}}
\newcommand{\bea}{\begin{eqnarray}}
\newcommand{\eea}{\end{eqnarray}}
\newcommand{\rf}[1]{(\ref{#1})}
\newcommand{\nn}{\nonumber\\}

\def\ha{(\widehat{k}_a){}}

\def\kad{(k_a^{(d)}){}}
\def\kat{(k_a^{(3)}){}}
\def\kaf{(k_a^{(5)}){}}

\begin{document}

\title{Neutral Mesons and Nonminimal CPT Violation}

\author{Benjamin R.\ Edwards}

\affiliation{Physics Department, Indiana University, 
Bloomington, Indiana 47405, USA}

\date{\today}

\begin{abstract}
Minimal CPT violation has been constrained 
using observations of neutral-meson oscillations.
Violation of CPT symmetry arising from nonminimal
operators in the Lagrange density can also occur.
A general approach
using scalar effective field theory
is presented and used to infer
the effects of nonminimal CPT violation
on neutral-meson oscillations.
\end{abstract}

\maketitle

{\it Talk presented at the 2019 Meeting of the Division of Particles and Fields of the American Physical Society (DPF2019), July 29 - August 2, 2019, Northeastern University, Boston, C1907293.}

\section{Introduction}

Both the Standard Model of Particle Physics
and General Relativity have been very successful
in describing much of the physics we have observed.
However,
it is also clear that they are in need of some modifications.
A breaking of Lorentz symmetry is one such candidate
that may occur in theories
combining quantum principles with gravity.
This has been shown to be possible,
for example,
in the context of strings \cite{KP}.
Any modifications should be tiny,
however,
in low-energy regimes
and therefore can be formulated using the tools
of effective field theory.
The Standard-Model Extension(SME) \cite{CK}\cite{AK04} is based on these ideas,
and is a framework for searches for Lorentz violation.
In addition,
a realistic effective field theory
that includes violations of CPT symmetry
necessarily violates Lorentz symmetry \cite{G}.
For this reason, 
the SME also provides a general framework
for studying CPT violation.

Constructed by adding terms at the level
of the Lagrange density that violate Lorentz symmetry,
the SME contains coefficients,
which are Lorentz scalars,
that control the size of the Lorentz-violating effect.
These are called the coefficients for Lorentz violation
and are the targets of experiments.
Some of the results of these investigations
are compiled in Ref.\ \cite{data}.
Limits of the SME including only
those terms with mass dimension
$d=3$ or $4$ make up what is referred to as
the minimal SME.
The rest of the framework including
terms with mass dimension $d > 4$
is referred to as the nonminimal SME.

CPT violation arising from
the minimal SME \cite{M1, M2, M3, M4} has been constrained in
experimental studies of neutral-meson oscillations \cite{M5, M6, M7, M8, M9, M10}.
Results are conventionally reported in a special frame
called the Sun-centered frame \cite{BKLRM},
with coordinate axes $T, X, Y, Z$,
and will also be used here.
The focus of what follows is to
obtain the effects of violations of CPT symmetry
on neutral-meson oscillations, 
including nonminimal terms.
The following material is based 
on the work found in Ref.\ \cite{EK19}.

\section{Formalism}

The formalism describing neutral-meson oscillations
involves a two-component state vector $\Ps$,
\beq
\Ps =
\left(\begin{array}{c}
\po \\ \opo
\end{array}\right)
\eeq
with $\po = \{ K^0, D^0 , B^0_d , B^0_s \}$
denoting one of the 4 neutral mesons.
The time evolution is governed by a
$2\times 2$ effective hamiltonian 
denoted by $\La$ and is determined by
$i\prt_t \Ps = \La \Ps$.
The propagating states
are the eigenstates of $\La$ written,
\bea
\ket{P_a(t)} &=& \exp (-i\la_a t) \ket{P_a},
\nonumber\\
\ket{P_b(t)} &=& \exp (-i\la_b t) \ket{P_b},
\label{timevol}
\eea
with eigenvalues
$\la_{a,b} \equiv m_{a , b} - \half i \ga_{a,b}$
where $m_{a,b}$
are the physical masses
 and $\ga_{a,b}$ are the decay rates.

The hamiltonian can be parametrized
in the following phase independent way \cite{M3}
\beq
\La =
\half \De\la
\left(\begin{array}{cc}
U + \xi
&
VW^{-1}
\\
& \\
VW
&
U - \xi
\end{array}\right),
\label{uvwx}
\eeq
where
$\De \la \equiv  \la_a - \la_b$.

The trace of $\La$ is tr$~\La = \la_a + \la_b$
and its determinant is $\det \La = \la_a \la_b$,
which imposes the conditions
\beq
U \equiv \frac{\text{tr }\La}{\De\la}, \quad
V \equiv \sqrt{1 - \xi^2}
\label{uvdef}
\eeq
on the complex parameters $U$ and $V$.
The independent parameters in Eq.\ \rf{uvwx}
can then be taken as
$W = w \exp (i\om)$,
and
$\xi = \Re\xi + i \Im \xi$.
The argument $\om$ of $W$ is physically irrelevant
and changes under phase redefinitions of the meson states.
The real modulus $w$ of $W$ controls T violation,
with $w = 1$ iff T is preserved.
The two real numbers $\Re\xi$ and $\Im\xi$ control CPT violation,
and both vanish iff CPT is preserved.
The relations between $w$, $\xi$
and the components of $\La$ are
\beq
w = \sqrt{|\La_{21}/\La_{12}|},
\quad
\xi = \De\La/\De\la,
\label{wxiexpr}
\eeq
where $\De\La = \La_{11}-\La_{22}$.
The form of $\xi$
depends on the underlying theory
and in what follows we will
be interested in finding the
effect of CPT violation on $\xi$,
including effects arising from terms in the nonminimal SME.

\section{scalar effective field theory}

Recent developments in the nonminimal
sector of the SME allow us to construct
the nonminimal terms in
both QED and QCD \cite{KZ},
however techniques have not been
developed to handle all nonminimal terms.
To simplify matters,
we view the meson as a point-particle
whose field operator is the complex scalar $\ph$
and focus on contributions to the meson propagator
due to nonminimal Lorentz violation
in a scalar effective field theory.
This allows us to infer the form of $\xi$
while leaving for later investigations
the proper match to the SME coefficients.

The general flavor $U(1)$ preserving
scalar effective field theory
incorporating violations of Lorentz symmetry
was studied in Ref.\ \cite{EK18}.
After incorporating flavor $U(1)$ breaking terms,
the contributions to the parameter for CPT violation
can be calculated.
Terms breaking $U(1)$,
however,
appear off diagonal in $\La$ and do not affect $\xi$.
In addition,
terms which preserve both
$U(1)$ and CPT contribute equally 
to the diagonal components of $\La$,
which again leave $\xi$ unaffected.
This means that only one class of terms contribute to $\xi$,
and these are the $U(1)$ preserving,
CPT-odd coefficients contained in
\beq
\cL \supset \half \big(i \ph^\dagger\ha^\mu\prt_\mu\ph 
+ \text{h.c.}\big) .
\eeq
The object $\ha^\mu$
can be expanded in momentum space as
$\ha^\mu =
\sum_{d=3}(k_a^{(d)})^{\mu{\al_1}\cdots{\al_{d-3}}}
p_{\al_1}p_{\al_2}\cdots p_{\al_{d-3}}$
with the sum running over odd $d$.

Note that an index indicating the meson species
has been suppressed for simplicity,
but that the coefficients can depend on the meson species as well.
The $\kad^{\al_1 \cdots \al_{d-2}}$ are coefficients for Lorentz violation 
in the scalar effective field theory,
and control the size of the Lorentz-violating effect.
They can be taken as symmetric and traceless,
and therefore have $(d-1)^2$ independent components.

\section{CPT-violating effects on $\xi$}

In order to calculate the contributions to the parameter $\xi$
arising from the coefficients for Lorentz violation,
it is convenient to focus on contributions
to $\xi^{(d)}$ arising from a single mass dimension $d$ term.
The full expression is then obtained by summing over $d$.
The hamiltonian can then be found directly
from the Lagrange density,
and the difference in diagonal terms
is proportional to $\xi$.
It can be shown that
\beq
\xi^{(d)} 
= \frac{1}{E_0\De\la}\kad^{\al_1 \cdots \al_{d-2}} p_{\al_1} p_{\al_{d-2}},
\eeq
for a meson with energy $E_0$.

The formalism introduced in the first section
is valid in the rest frame of the meson.
This means that in order to describe mesons not at rest
with components reported in the Sun-centered frame,
proper use of both observer 
and particle Lorentz transformations will be required.
Starting in the rest frame of a meson,
\beq
\xi^{(d)}
= \frac{m^{d-3}}{\De\la}
\kad_{0\cdots 0}^{\text{lab}}
\quad
\text{(rest frame of meson in laboratory)},
\eeq
where $m$ is the meson mass.

A particle boost is then employed to give
the meson a velocity $\be$ in the laboratory.
Under particle boosts,
the coefficient $k^{(d)}$
is a scalar,
while the meson 4-velocity undergoes 
$\be^\mu = (1,\vec{0}) \rightarrow \ga(1 , \vec{\be})$.
This produces an expression valid in the laboratory frame
\bea
\xi^{(d)}
&=& \frac{m^{d-3}\ga^{d-2}}{\De\la}
\sum_s {d-2 \choose s} \kad^{\text{lab}}_{0\cdots 0 j_1 \cdots j_s}\be^{j_1}\cdots \be^{j_s}
\quad
\text{(laboratory frame)},
\label{lab}
\eea

The components of the coefficients
in the laboratory frame,
with coordinate axes $j=$ $x$, $y$, $z$, and $t$
 are then related to the components in the
Sun-centered frame by an instantaneous observer rotation
denoted by $R$,
and the transformation is accomplished by
$\kad^{\text{lab}}_{0\cdots 0 j_1 \cdots j_s}
= \kad_{T\cdots T J_1 \cdots J_s}R^{J_1}_{j_1} \cdots R^{J_s}_{j_s}$.

Performing this rotation,
we arrive at the final expression
\beq
\xi^{(d)} = \frac{m^{d-3}\ga^{d-2}}{\De\la}
\sum_s {d-2 \choose s}\be^s
\kad_{T\cdots T J_1 \cdots J_s}\hat{\be}^J_1\cdots \hat{\be}^{J_s}
\label{xi}
\eeq
where the unit meson velocity in the Sun-centered frame
appearing in the expression above is related to the unit
meson velocity as measured in the laboratory by
\bea
\hat{\be}^\prime{}^X 
&=& (\hat{\be}^x \cos\ch  + \hat{\be}^z \sin\ch )
\cos{\Om T_\oplus}
- \hat{\be}^y \sin{\Om T_\oplus} \nn
\hat{\be}^\prime{}^Y
&=&
\hat{\be}^y \cos{\Om T_\oplus}
+ (\hat{\be}^x \cos\ch + \hat{\be}^z \sin\ch )
S_{\Om T_\oplus} \nn
\hat{\be}^\prime{}^Z
&=&
\hat{\be}^z \cos\ch - \hat{\be}^x \sin\ch.
\eea
The parameter $\ch$
is the angle between the local $z$-axis
in the laboratory and the $Z$-axis
of the Sun-centered frame.
The parameter $\Om$ is the sidereal frequency of the Earth
and is approximately
$\Om \approx 2\pi / (23 \text{ h } 56 \text{ min})$.
$T_\oplus$ is the local sidereal time.

\section{Inferred constraints}

The minimal $d=3$
type coefficients have been constrained
across all four meson species \cite{data}.
A match to the coefficients for Lorentz violation
in the scalar effective field theory
and the CPT-odd quark coefficients of the SME
can be found by comparing the known form
of the parameter $\xi$ arising from the minimal SME
to that arising from the scalar field theory above.
The match is found to be $\kat^\mu \approx 2\De a^\mu$.

Although the match for higher mass dimension is unknown at present,
constraints on the $d=5$
coefficients can be inferred from the published results
on the minimal SME coefficients,
taking care to address some subtle points.
First,
Eq.\ (\ref{xi}) includes trace terms.
By imposing the four conditions,
\beq
\kaf_{\mu TT} - \kaf_{\mu XX} - \kaf_{\mu YY} - \kaf_{\mu ZZ} = 0,
\eeq
we chose to eliminate the components of the type $(k^{(5)}_a)_{\mu ZZ}$,
satisfying the traceless condition.
Second,
the effects of minimal CPT violation
on $\xi$
only include variations in sidereal time in the first harmonic,
while at $d=5$
these variations include the second and third harmonics as well.
These harmonics accompany components of the type $\kaf_{\mu JT}$
and $\kaf_{\mu JK}$.
For this reason,
we had to assume only the components $\kaf_{\mu TT} \ne 0$.
Finally,
the velocity dependence at $d=5$
is more complicated than for the minimal coefficients already studied,
so we had to assume conservative values for these factors as well.

Constraints on the $TTT$ and $TTJ$ components of the $d=5$
coefficients are on the order of $10^{-18}$ $\rm{GeV}^{-1}$.
A full table of all results can be found in Ref.\  \cite{EK19}.
These constraints show that meson experiments are
competitive with other experimental methods \cite{exps}
sensitive to the CPT-odd quark coefficients of the SME.

\section*{Acknowledgments}
This work was supported in part by the
United States Department of Energy
under grant number DE-SC0010120
and by the 
Indiana University Center for Spacetime Symmetries.


\begin{thebibliography}{x}

\bibitem{KP}
V.A.\ Kosteleck\'y and R.\ Potting,
Nucl.\ Phys.\ B {\bf 359},
545 (1991). 

\bibitem{CK}
D.\ Colladay and V.A.\ Kosteleck\'y,
Phys.\ Rev.\ D {\bf 55},
6760 (1997);
Phys.\ Rev.\ D {\bf 58},
116002 (1998).

\bibitem{AK04}
V.A.\ Kosteleck\'y,
Phys.\ Rev.\ D {\bf 69},
105009 (2004).

\bibitem{G}
O.W.\ Greenberg,
Phys.\ Rev.\ Lett.\ {\bf 89},
231602 (2002).

\bibitem{data}
V.A.\ Kosteleck\'y and N.\ Russell,
{\it Data Tables for Lorentz and CPT Violation},
Rev.\ Mod.\ Phys. {\bf 83},
11 (2011),
2019 edition arXiv:0801.0287v12

\bibitem{M1}
V.A.\ Kosteleck\'y,
Phys.\ Rev.\ Lett.\ {\bf 80},
1818 (1998).

\bibitem{M2}
V.A.\ Kosteleck\'y,
Phys.\ Rev.\ D {\bf 61},
016002 (2000).

\bibitem{M3}
V.A.\ Kosteleck\'y,
Phys.\ Rev.\ D {\bf 64},
076001 (2001).

\bibitem{M4}
V.A.\ Kosteleck\'y and R.J.\ Van Kooten,
Phys.\ Rev.\ D {\bf 82},
101702(R) (2010).

\bibitem{M5}
KTeV Collaboration, H. Nguyen,
 hep-ex/0112046. 
 
\bibitem{M6}
FOCUS Collaboration,
J.\ Link {\it{et al.}}, 
Phys.\ Lett.\ B {\bf 556}, 
7 (2003).

\bibitem{M7}
BaBar Collaboration,
B.\ Aubert {\it{et al.}}, 
Phys.\ Rev.\ Lett.\ {\bf 100}, 
131802 (2008).

\bibitem{M8}
KLOE Collaboration,
D.\ Babusci {\it{et al.}}, 
Phys.\ Lett.\ B {\bf 730}, 
89 (2014);
A.\ Di Domenico,
Found.\ Phys.\ {\bf 40},
852 (2010).

\bibitem{M9}
V.M.\ Abazov {\it{et al.}},
Phys.\ Rev.\ Lett.\ {\bf 115}, 
161601 (2015).

\bibitem{M10}
LHCb Collaboration,
R.\ Aaij {\it{et al.}}, 
Phys.\ Rev.\ Lett.\ {\bf 100}, 
241601 (2016).

\bibitem{BKLRM}
R.\ Bluhm, V.A.\ Kosteleck\'y, C.D.\ Lane, N.\ Russell,
Phys.\ Rev.\ D {\bf 68},
125008 (2003);
Phys.\ Rev.\ D {\bf 88},
090801 (2002);
V.A.\ Kosteleck\'y and M.\ Mewes,
Phys.\ Rev.\ D {\bf 66},
056005 (2002).

\bibitem{EK19}
B.R.\ Edwards and V.A.\ Kosteleck\'y,
Phys.\ Lett.\ B {\bf 795},
620 (2019).

\bibitem{KZ}
V.A.\ Kosteleck\'y and Z.\ Li, 
Phys.\ Rev.\ D {\bf 99},
056016 (2019). 

\bibitem{EK18}
B.R.\ Edwards and V.A.\ Kosteleck\'y,
Phys.\ Lett.\ B {\bf 786},
319 (2018).

\bibitem{exps}
V.A.\ Kosteleck\'y, E.\ Lunghi, N.\ Sherrill, and A.\ Vieira,
in preparation;
V.A.\ Kosteleck\'y, E.\ Lunghi, and A.\ Vieira, 
Phys.\ Lett.\ B {\bf 769},
272 (2017);
M.\ Berger, V.A.\ Kosteleck\'y, and Z.\ Liu,
Phys.\ Rev.\ D {\bf 93},
036005 (2016). 

\end{thebibliography}
\end{document}